\documentclass{aa}

\usepackage{graphics}

\newcommand{\bp}{$\beta\:$Pictoris}

\begin{document}
\thesaurus{07(03.13.4; 05.15.1; 07.03.1; 08.16.2)}

\title{Photometric stellar variation due to extra-solar comets.}

\author{
A. Lecavelier des Etangs \inst{1}
\and A. Vidal-Madjar \inst{1}
\and R. Ferlet \inst{1}
}
        
\offprints{A. Lecavelier des Etangs}
\institute{
Institut d'Astrophysique de Paris, CNRS, 98bis Boulevard Arago, 
F-75014 Paris, France
}

\date{Received / Accepted}

\titlerunning{}
\authorrunning{}

\maketitle

\begin{abstract}

We performed numerical simulations of stellar occultations by extra-solar 
cometary tails. We find that extra-solar comets can be detected by the 
apparent photometric variations of the central stars. 
In most cases, the light curve shows a very peculiar
``rounded triangular'' shape. However, in some other cases, the
curve can mimic a planetary occultation.
Photometric variations due to comet occultations
are mainly achromatic. Nevertheless, if comets with small 
periastrons have smaller particles, these occultations could
be chromatic with a larger extinction in the blue by few percents. 

We also estimate the number of detections expected
in a large photometric survey at high accuracy. 
By the observation of several tens of thousand of stars,
it should be possible to detect several hundreds of occultation
per year.
We thus conclude that a spatial photometric survey would
detect a large number of extra-solar comets. 
This would allow to explore the time evolution of cometary activity,  
and consequently would probe structure and evolution of extra-solar 
planetary systems.


\end{abstract}

\section{Introduction}

The search for extra-solar planets through photometric variation is
a well-known problem analyzed in many aspects for several years 
(Schneider et al. 1990; Schneider 1996).
However, planets are not the only objects detectable by their effect
on stars' brightness, comets can also induce photometric
variations.
In our study of \bp\ photometric variations (Lecavelier des Etangs
et al. 1995), we 
concluded that the variations observed on November 1981
could be due to the passage in front of the star 
of either a planet or a dusty cloud 
(Lecavelier des Etangs et al. 1997, Lamers et al. 1997). 
If the latter is the correct explanation, this cloud of dust must
be a cometary tail for two reasons. First, 
because the lifetime of the dust
in such a system is very short and 
one must find a way to produce the dust continuously.
Second, because the cloud shape is constrained by the 
observed light curve; the cloud
cannot be spherical, it should have a sharp edge 
in the part pointing toward the star and a huge cloud of dust 
in the opposite direction: exactly like a cometary tail (Lecavelier des Etangs,
1996).

In the case of the solar system, the observations of stellar occultation by comets
have been discussed by Combes et al. (1983).
Observations of extinction and polarization of star light 
by dust of cometary tails in the solar system have been published
(Dossin 1962, Ninkov 1994, Rosenbush et al. 1994).

But the solar system is only one particular planetary system at a given age.
It is known that cometary activity was formerly much more important,
and the well-known case of \bp\ shows that during the first
$10^8$~years, a planetary system is expected to show
large cometary activity (Ferlet \& Vidal-Madjar 1995, 
Vidal-Madjar et al. 1998).
Moreover, presence of comets around stars can be considered
as an indirect signature for the presence of gravitational
perturbations, and possibly caused by planets.

In this paper, we deal with the important possibility of detecting
cometary activity from a photometric survey.
We take the COROT space mission as an example of what will be achievable
in the very near future (Baglin et al. 1997). 
COROT, which primary aim is stellar seismology, will be launched 
in early 2002. It will allow a survey of
about 30~000 stars with a photometric accuracy of few $10^{-4}$
during several months.
Here, we predict the probabilities of detecting 
comets using such a photometric survey.

In Sect~\ref{A model of cometary occultation}, we describe the model 
of the cometary occultation, 
and then give the expected light curve in Sect~\ref{Light curves section}.
Estimates of the number of comets which could be detected are given 
in Sect~\ref{Probability of detection}. The conclusion is in
Sect~\ref{Conclusion}.

\section{A model of cometary occultation}
\label{A model of cometary occultation}

As pointed out by Lamers et al.(1997), a geometric distribution of the dust 
must be assumed in order to evaluate the 
photometric stellar variation due to extra-solar comets.
Then, by taking into account the optical properties of the cometary grains, 
the photometric variation can be estimated.

\subsection{Distribution of dust in a cometary tail}
\label{Distribution of dust in a cometary tail}

The evaluation of the distribution of the dust in a cometary tail
can be made through particle simulation. 
The input parameters are the comet orbit, 
the dust production rate, the ejection velocity and the size distribution.
We assume a size distribution $dn(s)$ of the form
\begin{equation}
dn(s)=\frac{(1-s_0/s)^m}{s^n}
\label{dn(s)}
\end{equation}
as observed in the solar system, 
where $s$ is the dust size. We take $s_0=0.1\mu$m,
$n=4.2$, $m=n(s_p-s_0)/s_0$, and $s_p=0.5\mu$m (Hanner 1983,
Newburn \& Spinrad 1985).
This distribution starts at $s_0=0.1\mu$m with $dn(s_0)=0$, peaks at
$s_p=0.5\mu$m, and has a tail similar to a $s^{-n}$ distribution
for large sizes.

Dust sensitivity to radiation pressure is given by
the $\beta$ ratio of the radiation force to the gravitational force. 
We take 
\begin{equation}
\beta=0.2 \left( \frac{L_*/M_*}{L_{\sun}/M_{\sun}}\right)
\left( \frac{s}{1\mu {\rm m}}\right)^{-1}
\end{equation}
where $L_*/M_*$ is the luminosity-mass ratio of the star.
So the orbits of the small grains are more affected by radiation pressure
than those of large grains.
This is a very good approximation for particles larger than
$0.1\mu$m of any realistic composition and for the solar spectrum
(Burns et al. 1979).

The dust production rate $P$ is assumed to vary with $r$, the distance to the
star, and is taken to be
\begin{equation}
P=P_{0} \left(\frac{r}{r_0} \right) ^{-2}
\end{equation}
(see, for example, 
A'Hearn et al., 1995; Weaver et al., 1997, Schleicher 1998).
The dust production is taken to be zero beyond 3~AU for a solar
luminosity.

As soon as a grain is produced from a comet nucleus, we assume that
it is ejected from the parent body with a velocity $v_{\rm eject}$ 
in an arbitrary direction.
The grain then follows a trajectory defined by gravitation and radiation forces. 
The ejection velocity depends upon the particle size.
We take $v_{\rm eject} = \sqrt{\beta}/(A+B\sqrt{\beta})$
(Sekanina \& Larson 1984),
which approximates the results of Probstein's (1969) two-phase dusty-gas
dynamics for the acceleration by the expanding gas within tens of kilometers
from the nucleus. The coefficient $A$ and $B$ depend upon many parameters 
such as the thermal velocity of the expanding gas.
We used $A=B=1$~s~km$^{-1}$ which is a good approximation
of different values measured for the comets of the solar system
(Sekanina \& Larson 1984; Sekanina 1998).
This gives an ejection velocity of $\approx 585$km~s$^{-1}$ for $s=0.1\mu$m.
The ejection velocity is smaller for larger grains which have smaller
cross section area to mass ratio ($v_{\rm eject}(10\mu{\rm m}) \approx
124$km~s$^{-1}$). 
We checked that any other realistic values
for $A$ and $B$ give similar light curves within few percents.

\subsection{Stellar parameter}

The simulations have been performed with the
mass, luminosity and radius of the central star set to 
solar values ($M_*$, $L_*$ and $R_*$). Simulation with other
parameters could be possible. A larger mass for the central star
would induce a shorter time scale; a larger luminosity would
increase the effect of the radiation pressure on grains;
a larger radius would decrease the relative extinction 
(Eq.~\ref{Fext/F*}).
However, the conclusions do not strongly depend on these stellar properties.
For instance, adopting the properties of an A5V star does not change the shape
of the light curves. By scaling the production rate
to the star luminosity, we found a quantitative change 
by less than a factor of two.

\subsection{Grain properties}
\label{Grain properties}

\subsubsection{Extinction}

The extinction cross-section of a dust grain is 
$Q_{ext}(s,\lambda) \pi s^2$, where the extinction efficiency,
$Q_{ext}$ is slightly dependent on the particle size ($s$) and radiation 
wavelength ($\lambda$). 
If $Q_{sca}$ is the scattering efficiency (see Sect.~\ref{Scattering})
and $Q_{abs}$ is absorption efficiency, 
$Q_{ext}(s, \lambda) = Q_{sca} + Q_{abs}$. We take 
$Q_{abs}=1$ if $s\ge\lambda$, and
$Q_{abs}= s /\lambda$ if $s <\lambda$, which
is a good approximation for optical wavelengths and grains 
larger than 0.1$\mu$m (Draine \& Lee 1984).

The total extinction is hence calculated by adding
the extinction due to all particles in the line of sight to the star.
The optical depth $\tau$ due to the dust is
\begin{equation}
\tau=\frac{ \sum_{\rm part.} N_{\rm grain/part.} Q_{ext}(s,\lambda) \pi s^2}
          {S}
\label{tau_ext}
\end{equation}
where $S$ is the projected area of the line of sight.
In the simulation, one particle represents an optically thin cloud of 
several dust grains. The number of physical grains per particle in the 
simulation, $N_{\rm grain/part.}$, is
\begin{equation}
N_{\rm grain/part.}=\frac{3M_{dust}}{4\pi \rho N_{\rm part.}}
   \frac{\int dn(s)}{\int s^3 dn(s)}
\label{Ngp}
\end{equation}
where $M_{dust}$ is the total dust mass, $\rho$ is the dust density, and
$N_{\rm part.}$ is the number of particles in the simulation.
$N_{\rm part.}$ is set to a few $10^4$ in order to
keep reasonable computing time.

Because the cometary cloud is optically
thick but its size is smaller than the size of the star, 
we mapped the stellar surface
through a set of cells in polar coordinates. For each cell $i$, we calculate
the optical depth $\tau_i$ due to the particles within 
this cell of area $S_i$ ($\sum_i S_i = \pi R_*^2$).
The ratio of the flux observed through the cloud ($F_{ext}$) to the initial
stellar flux ($F_*$) is
\begin{equation}
\frac{F_{ext}}{F_*} = \sum_i \frac{S_i e^{-\left(\tau_i \right)}}
{\pi R_*^2} 
\label{Fext/F*}
\end{equation}

The number of cells is the best compromise between the spatial resolution and
the number of particles in the simulation. We take into account that
the maximum contribution to $\tau_i$ by each
particle must be $<10^{-1}$ in order to achieve an accurate result
in spite of the quantization of the extinction.
$S_i$ and $N_{\rm part}$ are thus constrained by
\begin{equation}
\frac{S_i N_{\rm part}}{10} \ge\frac{3M_{dust}Q_{ext}}{4\rho}
   \frac{\int s^2 dn(s)}{\int s^3 dn(s)} 
\end{equation}

The limb darkening is not taken into account in this work, 
because its effect is negligible.

\subsubsection{Scattering}
\label{Scattering}

As already pointed out by Lamers et al. (1997),
the main part of an occulting dust cloud is seen 
through a very small scattering angle. Thus, the total
star light forward-scattered to the observer can be large
because the phase function is strongly peaked 
to small angles for which the diffraction 
has the dominant contribution.
The phase function for the diffraction is
\footnote{The Eq.~12 in Lamers et al. (1997) has a typo-mistake: 
$\tilde x$ in the first fraction must be $\tilde x^2$. }
\begin{equation}
P(\theta, \tilde x)=\frac{\tilde x^2 \left( 1 + \cos^2 \theta \right) }{4 \pi}
\left(\frac{J_1(\tilde x \sin \theta)}{\tilde x \sin \theta} \right)^2
\end{equation}
where $\tilde x=2 \pi s / \lambda$.
The phase function has been normalized by $\int P(\theta)d\Omega =1$
and depends upon the wavelength $\lambda$.
Examples of such phase functions are shown in Lamers et al. (1997).

The scattered light ($F_{sca}$) is evaluated by adding the contribution 
of each particle in the simulation.
\begin{equation}
\frac{F_{scat}}{F_*}=
N_{\rm grain/part.} \sum _{\rm part.} 
\frac{Q_{sca} \pi s^2 P(\theta)}{r^2} \,\,e^{-(\tau _{in}+\tau _{out})}.
\end{equation}
$\tau _{in}$ is the total extinction along the path from
the star to the scattering grain and $\tau _{out}$ 
is from the grain to the observer.
The grain is at distance $r$ from the star.
For $s\ge\lambda$, the scattering efficiency 
$Q_{sca}$ is assumed to be the diffraction
efficiency for large grains: $Q_{sca}=Q_D=1$ 
(Pollack \& Cuzzi, 1980).
For small particles ($s < \lambda$), we used the basic approximation 
$Q_{sca}=(s /\lambda)^4$ (van de Hulst, 1957). The result is very insensitive 
to this last assumption 
because forward scattering at small angle is largely 
dominated by diffraction on particles larger than the wavelength.
Because each cloud represented by a particle is very thin, and the total
extinction is small ($\tau_{in}$, $\tau_{out} <10^{-1}$), 
we assumed single scattering (no source function). 

In fact, for very peaked forward-scattering function on very small
scattering angle, the finite size of the star must be taken into account
(especially when the dust cloud is seen superimposed on the star surface).
Hence, we mapped the stellar surface by small arcs centered on the 
dust particle. We then calculate the total scattering
by adding the contribution of each arc.

\section{Light curves}
\label{Light curves section}

The light observed at a given time is the sum
of two opposite effects: the increase of the brightness through the scattering 
by particles at small angle from the line of sight and 
the decrease of the brightness through the extinction (including
absorption) by particles on the line of sight. 

\subsection{First order estimations}

\subsubsection{Extinction}

From Eqs.~\ref{tau_ext} and~\ref{Ngp}, in the case 
$s \ga \lambda$ ($Q_{ext}=2$), and
neglecting the inner extinction, we find an upper limit for the
extinction:

\begin{equation}
\left(\frac{F_{ext} }{F_*}\right) _{\rm max} 
= \exp\left(
   -\frac{6M_{dust} \int s^2 dn(s) }{4\pi \rho R_*^2 \int s^3 dn(s)}
     \right) 
\end{equation}

For the particle size distribution given by Eq.~\ref{dn(s)}
(Sect.~\ref{Distribution of dust in a cometary tail}),
we have
$(\int s^2 dn(s))/( \int s^3 dn(s))= 0.15 \mu^{-1}$. 
Adopting a dust production of $10^6$~kg~s$^{-1}$ during 100 days,
we find $M_{dust}\approx 10^{13}$~kg. 
Hence $(\Delta F / F_*)_{\rm max} \approx 10^{-3} $. 
This order of magnitude is close to the result of  more
elaborated calculations (Sect.~\ref{Light curves}).
We can already conclude that 
this variation could be detected by a space photometric survey.

We note that the photometric variations of $\beta$~Pic in 1981 was $\sim
5\cdot 10^{-2}$. The above calculation gives an estimate of the
lower limit of the needed production rate : $P \ga 5\cdot
10^7$~kg~s$^{-1}$. This is consistent with $P \ga 10^8$~kg~s$^{-1}$
obtained from the estimation of Lamers et al. (1997): 
$P/ \Delta v > 10^{8}$kg~km$^{-1}$ and with an assumed relative velocity 
between the dust and the nucleus of the comet $\Delta v\la 1$km~s$^{-1}$.

\subsubsection{Scattering}

If we take a comet at $r=1$~AU from the star and
an impact parameter on the line of sight of 
one half the stellar radius,
then $\theta\approx 7\arcmin $. For the particle size distribution
given by Eq.~\ref{dn(s)}, the phase function is 
$\tilde P(\theta\approx 7\arcmin)\approx 700$. 
Therefore,
\begin{equation}
\frac{F_{scat}}{F_*} 
\approx
 \frac{3M_{dust}  \tilde P(\theta)  } {4 \rho r^2 }
 \frac{\int s^2 dn(s)}{\int s^3 dn(s)}
\approx
 4\cdot 10^{-5}
\end{equation}

As a first conclusion, it is clear that both extinction and scattered light 
can be detected by very precise photometric survey, although scattering gives
photometric variations an order of magnitude smaller than extinction.
Extinction appears to be 
the major process observed when the comet is passing
in front of the star. This event can thus be called an occultation.

\subsection{Light curves}
\label{Light curves}

Taking into account the dust spatial distribution and the extinction
within the tail, we calculated the light curves of 
a set of cometary occultations. From a given comet orbit,
and dust production rate, 
we compute the full motion of each dust particle;
the result is the variation of the star light as a function 
of time.

\begin{figure}
\resizebox{\hsize}{!}{\includegraphics{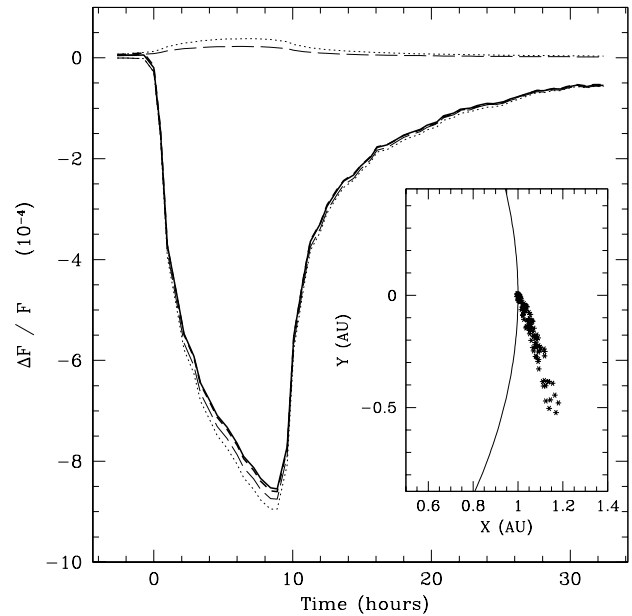}}
\caption[]{Plot of the photometric variations during a cometary occultation 
in red ($\lambda\sim 8000$~\AA, thick solid line) and in blue ($\lambda\sim
4000$~\AA, thick short-dashed line). 
The insert is a view from the top when the 
comet is crossing the line of sight ($Y=0$)
at the periastron. The production rate is $2\cdot 10^6$~kg~s$^{-1}$
at 1~AU. The scattered light is given by the top thin lines, the 
extinction by the bottom thin lines
(long-dashed line for $\lambda\sim 8000$~\AA; 
dotted line for $\lambda\sim 4000$~\AA). 
The total variation is plotted with the thick lines. 
This light curve presents the very specific "rounded
triangular" shape observed in the majority of simulations of cometary 
occultations. 
The difference between the variation in the blue and in the red 
is negligible and less than 1\%.}
\label{fig1}
\end{figure}

\begin{figure}
\resizebox{\hsize}{!}{\includegraphics{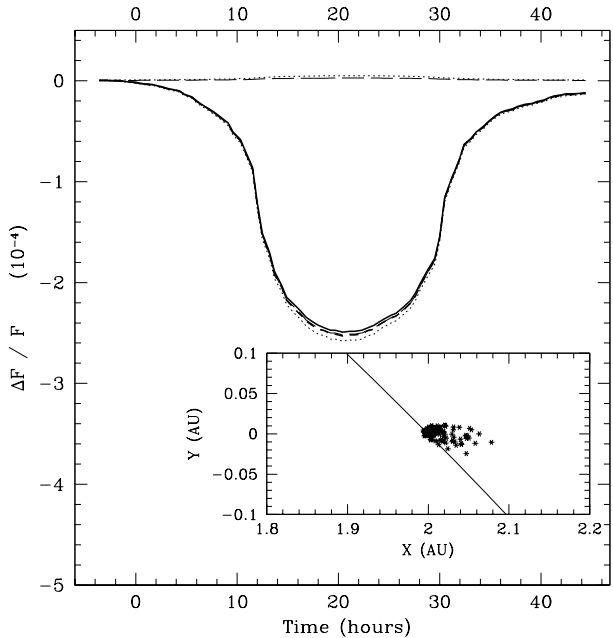}}
\caption[]{Same as Fig.1. Here all the parameters are the same except that
the longitude of the periastron is at 90$\degr$ from the line of sight. 
This gives a production rate of $5\cdot 10^5$~kg~s$^{-1}$ at 2~AU. 
The cometary tail is aligned with the line of sight; consequently the 
light curve presents a more symmetrical shape resembling
a planetary occultation.
The net color effect is hardly detectable and of the order of 1.5\%.}
\label{fig2}
\end{figure}

Two typical light curves resulting from the simulation are 
shown in Fig.~\ref{fig1} 
and~\ref{fig2}. The majority of cometary occultations gives light curves
with a very particular ``rounded triangular'' shape (Fig.~\ref{fig1}). 
This occurs when the dense cometary head first occults the star and 
gives a very fast and sharp brightness decrease.
It is then followed by the tail which gives an additional slow decrease.
Similarly, the subsequent brightness increase is also sharp when
the cometary head is going out of the occulting part.
Then the increase slows down and the brightness returns to the normal level 
when the cometary tail is less and less dense in front of the star.

\begin{figure}
\resizebox{\hsize}{!}{\includegraphics{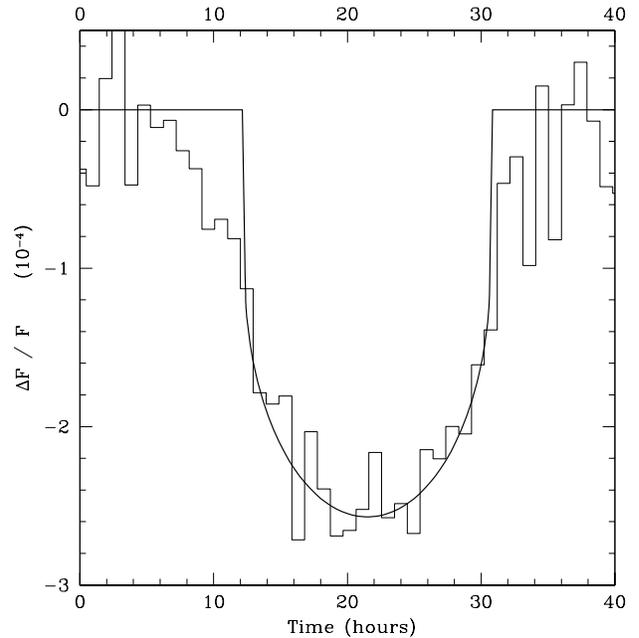}}
\caption[]{Plot of the photometric variations due to a comet 
similar to the comet of the Fig.2 (histogram). 
Here the result is given as a noisy observation
with a photometric measurement every hour and an accuracy 
of $3\sigma=10^{-4}$. For comparison, an occultation
by a planet with a radius of 9~000~km and orbiting at 2~AU
is plotted with limb darkening effect taken into account (thick line).
The detected variation is very similar to a planetary 
occultation, the main difference being extended wings in the cometary case.
}
\label{fig3}
\end{figure}

However, in some configurations, the tail can be aligned
with the line of sight. In these cases the light curves
are more symmetric (Fig.~\ref{fig2}).
They can mimic planetary occultations (Fig.~\ref{fig3}).
Because of the noise, it will be difficult from 
such observations to differentiate between a comet and a planet.

\subsection{Color signatures}

\begin{figure}
\resizebox{\hsize}{!}{\includegraphics{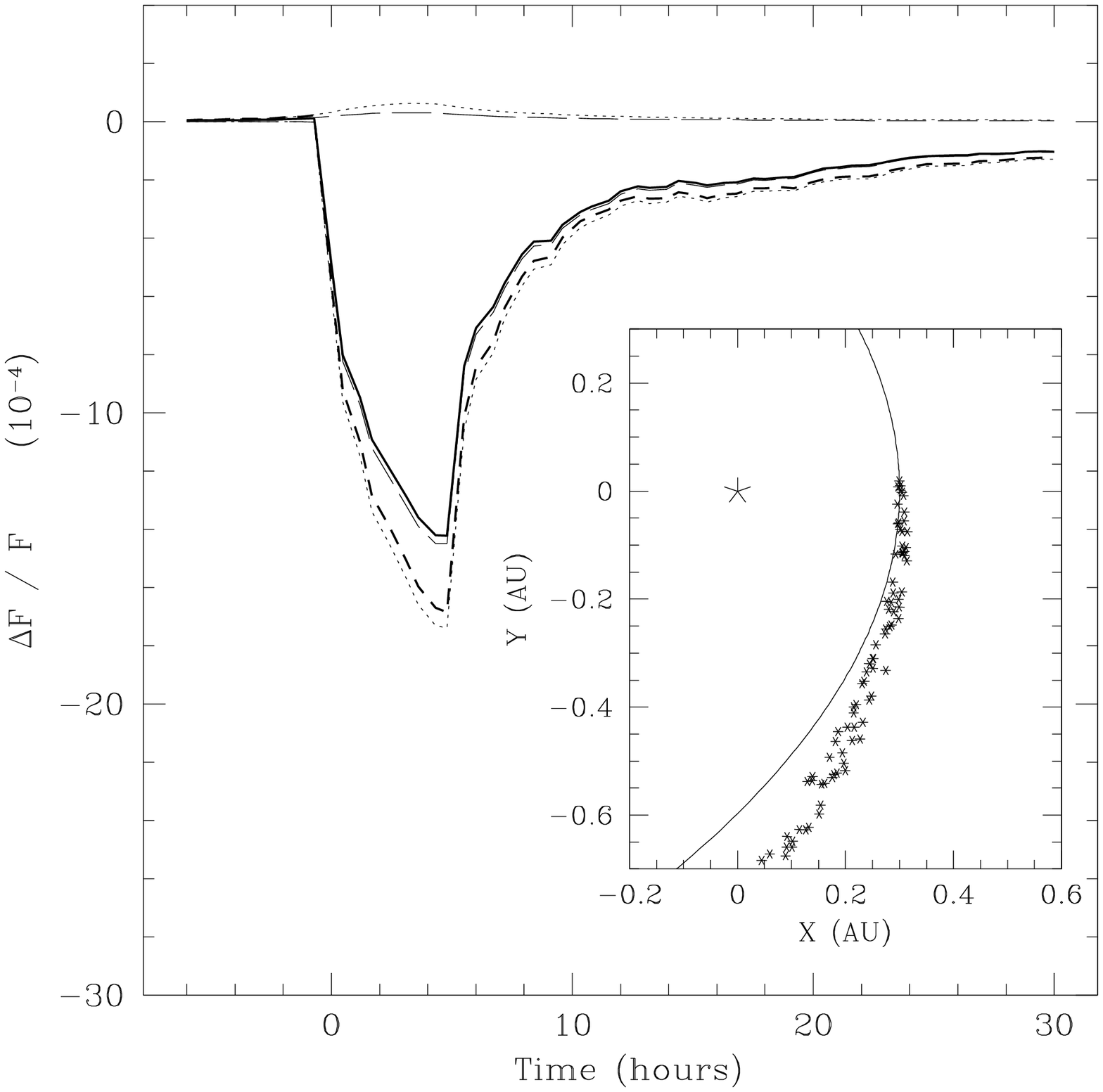}}
\caption[]{Same as Fig.1. Here all the parameters are the same except that
the periastron is at 0.3~AU and the particle size distribution is
taken with $a_p=0.25\mu$m.
The production rate corresponds to $2\cdot 10^7$~kg~s$^{-1}$ at 0.3~AU. 
Because there are more particles smaller than the wavelength, 
we see a dipper occultation in the blue by about 15\%. }
\label{fig4}
\end{figure}

As seen in Sect.~\ref{Grain properties}, the light variations 
show some color signature due to the optical properties of the grains.
Particles with a size smaller than the wavelength
are less efficient for extinction. Thus, extinction
is smaller at larger wavelengths. 
The forward scattering is more peaked to
the small angles at shorter wavelengths (Lamers et al. 1997).
The scattering is thus larger at smaller wavelengths when the
cometary cloud is occulting the star ({\it a contrario}, the scattering
is larger at larger wavelength when the comet cloud is seen far from the star;
but then the scattered light becomes negligible).
As a result, there is a balance between the additional scattered light 
and the extinction; both are larger at smaller wavelengths. 
It is difficult to predict what will be the color effect on cometary
occultation light curves.

In addition, most of the extinction and scattering 
due to the cometary dust is concentrated in the inner coma, where
the number of large particles is larger because smaller
particles are more efficiently ejected by radiation pressure
(Sect.~\ref{Distribution of dust in a cometary tail}).
As a consequence, in the optical, and with the size distribution assumed
in Sect~\ref{Distribution of dust in a cometary tail},
the light curves are barely dependent on the
wavelength (or color band). The difference 
between variations in blue and in red is smaller than 
few percents. This color signature would be very difficult to 
observe. This is beyond the today technical feasibility.

However, for comets at small distances from the Sun ($\la 0.5$~AU), 
its appears that the dust size distribution is peaked at smaller sizes 
(Newburn \& Spinrad 1985).
This decrease of $a_p$ with the heliocentric distance could be due to
particle fragmentation. Although it is difficult to guess 
the size distribution for extra-solar comets, it is easy to understand 
that if small particles are more numerous (with $s<0.5\mu$m), the 
occultation will show a larger color signature.
We checked that a comet with a periastron at 0.3~AU and $s_p=0.25\mu$m 
(Newburn \& Spinrad 1985)
gives larger extinction by about 15\% in the blue than in the red
(Fig.~\ref{fig4}).

\subsection{Conclusion}

Photometric variations due to the stars' occultation
by extra-solar comets could be detectable by photometric measurements
with an accuracy of $\sim 10^{-3}$ -- $10^{-4}$. In most cases, 
the particular "rounded triangular" shape of the 
light curve can be an easy diagnostic for the presence of a comet.
However, in some cases, the light curves can mimic an occultation by a more 
compact object like a classical extra-solar planet (Lecavelier des Etangs
et al. 1997). The color signatures of cometary occultations 
can be too small to avoid the confusion.
The detection of periodicity appears to be critical in the diagnostic of a
planetary occultation.
Alternatively, the polarization is another way to discriminate
the two {\it phenomena}, because the light
is scattered by the cometary dust roughly gathered in the same plane.
But, with a level of at most few percents in polarization of solely the
scattered light, this gives $<$ 0.01 \% polarization in total and would also 
be difficult to detect. In cases of planetary and cometary occultations,
spectroscopic follow-up observations should be planned to 
allow a better analysis of a suspected on-going detection.

\section{Probability of detection}
\label{Probability of detection}

\begin{figure}
\resizebox{\hsize}{!}{\includegraphics{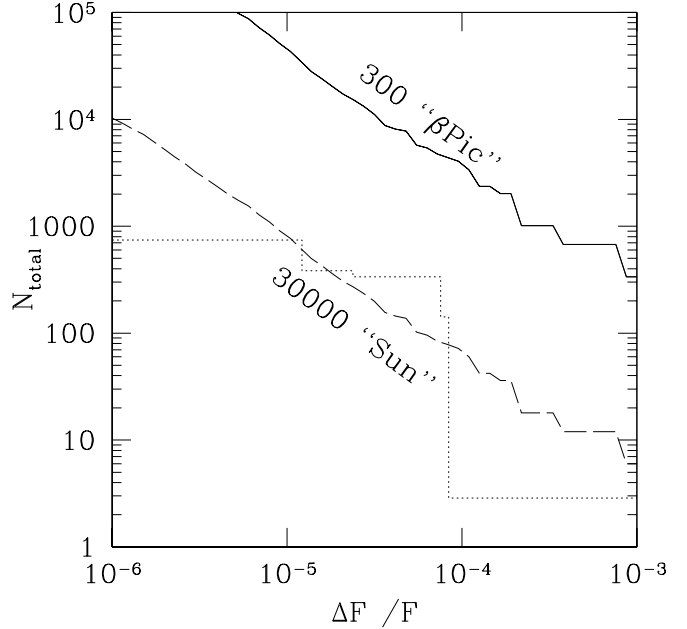}}
\caption[]{Plot of the number of detections of cometary occultations
as a function of the photometric accuracy. Two cases
are considered: the ``pessimistic case'' which is a survey of
$n_{star}=30000$~sun-like stars, during $T_{obs}=1$~year
($n_{comet}=100$, $P_0=10^3$~kg~s$^{-1}$), and the ``optimistic case''
where among the 30000 stars, 1\% ($n_{star}=300$) have a $\beta$~Pictoris-like
activity ($n_{comet}=100$, $P_0=10^6$~kg~s$^{-1}$).
With an accuracy of $10^{-4}$, about 10 to 10$^3$ detections
can be expected. The large difference between the two cases shows
that this kind of survey will also give information on the planetary
evolution.

For comparison, the dotted histogram gives the number of detection of
planets assuming that each star has a planetary system like the
solar system. With an accuracy larger than $10^{-4}$, this gives the
number of detection
of giant planets, and mainly Jupiter-like planets. With an accuracy better
than $10^{-4}$, Earth-like planets will be detected.
Each step in the histogram represents the possibility to detect
successively the Earth ($8\cdot 10^{-5}$), Venus ($7\cdot 10^{-5}$),
Mars ($2\cdot 10^{-5}$) and Mercury ($10^{-5}$). We see that accurate
photometric survey should detect more comets than planets.}
\label{ntotal}
\end{figure}

From the above modeling, we can evaluate the probability to detect a 
photometric stellar variation due to an extra-solar comet.
We assume a survey of a given number of stars at a given accuracy.
We carried out a large number
of simulations of cometary occultations in all directions. 

\subsection{Orbital parameters}

The characteristics of the comets are given by their distribution
in the parameter space. Concerning the orbital parameters,
the periastron density distribution is chosen to be the same as 
the one observed in the solar system:
$dn(q) \propto q^{0.3}dq$
with $q\in [0.1~$AU$, 2~$AU$]$ (A'Hearn et al., 1995). 
Many sun-grazing comets have been discovered by 
SOHO with perihelion $q<0.1$~AU (Kohl et al. 1997; see also 
the numerous IAU Circulars). But it is still difficult to infer
the distribution for these small periastrons. 
The above distribution is thus likely biased 
by an underestimate of the number of comets with small periastron,
because such comets are difficult to observe. These star-grazing
comets give a larger photometric variation. Therefore,
we possibly slightly underestimate the probabilities of detection.

We fixed all the apoastrons at 20~AU, 
the longitudes of periastron, the ascending nodes and the inclinations
are chosen randomly.

\subsection{Dust production rate}

One of the most important parameter is the dust production rate. 
Because, it dictates the
mass of dust in the tail, and constrains the detectability of the
comet. The amplitude of the photometric variation
is roughly proportional to that parameter. 

We consider that the dust production rate ($P$) is proportional 
to the area of the comet's nucleus, 
and that the distribution of the comets' radii ($R_{c}$) is similar to 
the distribution observed for
the comets, asteroids, and Kuiper belt objects of the solar system.
We assume that the number density of objects with radius 
in the range $R_{c}$ to $R_{c}+dR_{c}$ is 
$dn(R_{c})\propto R_{c}^{-\gamma}dR_{c}$.
$\gamma$ is a positive number, typically in the range 3 to 4 (Luu 1995). 
It is constrained to be $\sim$ 3 -- 3.5 for the comets' nuclei
observed with $R\in [0.1$km$, 100$km$]$ (Fern\'andez 1982,
Hughes \& Daniels 1982, Brandt et al. 1997)
The same distribution is 
consistent with the observation of the Kuiper belt objects:
$\gamma \sim 3$ with $R\in [100$~km$, 400$~km$]$ (Jewitt 1996), 
or with theoretical models for the formation of these objects:
$\gamma \sim 4$ for $R\la 200$~km (Kenyon \& Luu, 1998).
With the assumption that $P\propto R_c^2$, 
the number of comets with a production rate between 
$P$ and $P+dP$ is $dn(P)\propto P^{-(\gamma+1)/2}dP$.
We adopt $\gamma = 3.5$. Changing $\gamma$ to 3 or 4 would 
change the probability of detection by less than a factor of two.

Finally, this distribution is normalized by $n_{comet}$, the number of comets
per unit of time passing through the periastron
with a production rate larger than $P_0$. Hence, we have
$n_{comet}\propto \int_{P_0}^{\infty} dn(P)$. Thus,
\begin{equation}
 n_{comet} \propto P_0^{(1-\gamma)/2}.
\end{equation}

\subsection{Results}

Using a large number of various comets, 
we calculate the probability of detection at a given photometric accuracy. 
Then, the number of possible detections is simply
this probability multiplied by $n_{comet}$, the duration of the observation
$T_{obs}$ and the number of stars surveyed $n_{star}$.

We suppose that the time scale between each measurement
is small enough ($\la 1$~hour) that each variation 
above the detection limit will effectively be 
detected. As an example, we take  $T_{obs}=1$~year and
$n_{star}=30\,000$, which is the order of magnitude
for the future space mission COROT.

The expected number of detections is plotted in Fig.~\ref{ntotal}.
We considered two types of planetary systems. The first one is
similar to the solar system with $n_{comet}=100$ comets per year
with $P\ge P_0= 10^3 $~kg~s$^{-1}$ at $r_0=1$~AU. We see that few dozens of 
comets could be detected at an accuracy of $10^{-4}$. We consider it 
as the pessimistic case.

The second type of planetary system is supposed to be
a young planetary system with a large cometary activity
as during the youth of the solar system.
The typical example is the well-known star $\beta$~Pictoris
where comets' infalls are commonly observed through 
UV and optical spectroscopy 
(see e.g., Ferlet et al. 1987, Lagrange et al. 1988, Beust et al. 1990,
Vidal-Madjar et al. 1994, 1998).
For such a planetary system, $n_{comet}=100$ comets per year
with $P\ge P_0= 10^6 $~kg~s$^{-1}$ at $r_0=1$~AU
(Beust 1995, Beust et al. 1996).
Note that $\beta$~Pictoris is young but on the main sequence
(Crifo et al. 1997), its age is few 
percents of the age of the solar system.
Therefore, in a set of 30\,000 stars
there should be about $n_{star}\approx 300$ stars with about the same activity
as $\beta$~Pictoris.
Thus, few thousands comets could be expected with a survey at $10^{-4}$ 
accuracy. We consider it as the optimistic case.

The solar system is likely not exceptional. 
The bottom-line in Fig.~\ref{ntotal} is a good
estimate of the lower limit of the expected number of detections.
Younger stars may have a higher level of activity, with a larger number
of comets; but, although infrared excess have been observed around many main
sequence stars, $\beta$~Pictoris 
is certainly a very peculiar case (Vidal-Madjar et al. 1998).
Thus, the top-line in Fig.~\ref{ntotal} gives a good
estimate of the upper limit of the expected number of detections.

Note that the major assumption that the dust production rates in the solar system 
can be extrapolated to large values is realistic.
A large rate has effectively been observed in the recent comet Hale-Bopp 
where it reached of few times $10^{5}$~kg~s$^{-1}$ at about 1~AU
(Rauer H. et al., 1997, Schleicher et al., 1997, Senay et al. 1997,
Weaver et al., 1997).

It is very likely that in the near future 
a large number of extra-solar comets will be detected through occultations.

For comparison, we also evaluate the probability to detect  
planets assuming that each star has a planetary system like the
solar system. 
With an accuracy better than $10^{-2}$, Jupiter can be detected.
Below $10^{-3}$, other giant planets are also detectable. But
because of their large orbital periods,  
their contribution to the number of detections is smaller than 
the one from Jupiter.
With a total probability of $\sim 10^{-4}$ to detect a giant planet 
in one year, the survey of 30000 stars will give $\sim 3$ 
planets.
With an accuracy better
than $10^{-4}$, Earth-like planets start to be detectable.
Because of their smaller distance to the star, 
they have larger contribution to the probability of planet detection.
The Earth and Venus can be detected with a probability of
$\sim 10^{-2}$ in one year. Mars and Mercury, which are detectable
with an accuracy of $\sim 10^{-5}$, give a total probability
of $\sim 3\cdot 10^{-2}$. 
The comparison with the number of detection of comets shows that
accurate photometric surveys should detect more comets than planets.

\section{Conclusion}
\label{Conclusion}

We performed detailed 
numerical simulations of stars' occultations by extra-solar 
comets. We extracted the apparent photometric variations of the central 
stars due to these putative comets. 
We have shown that:

1) Extra-solar comets can be detected through
photometric variations due to occultation by 
dusty tails. In many cases, the light curve shows a very particular
``rounded triangular'' shape. However, in some remaining cases, the
curve can mimic a planetary occultation.

2) The photometric variations due to cometary occultations
are mainly achromatic. This property will allow to 
discriminate the occultations by comets from intrinsic stellar variations. 
However, the confusion with planetary occultations
cannot be efficiently removed by color measurements in the optical.

3) The number of detections which can be expected from a large photometric 
survey of several tens of thousand of stars at high accuracy ($10^{-4}$) 
is of the order of several hundreds of occultations per year.

These detections will allow to explore the evolution of
the cometary activity through the correlation with the stellar age.
Indeed, we know from the solar system exploration that the cometary activity 
significantly changed with time. 
Moreover, comets are believed to be the primitive bricks
of the planetary formation, and planets perturbation are
needed to push them toward the central star.
It is thus clear that the detection and analysis of the cometary 
activity around nearby stars will give important information on 
the structure and evolution of the planetary systems.

\begin{acknowledgements} 
We would like to express our gratitude to Alain Leger for fruitful discussions.
We also thank the referee Henny Lamers for his useful comments 
which improved the paper. \\
We warmly thank Remi Cabanac for his critical reading of the paper.
\end{acknowledgements}

\end{document}